\documentclass[useAMS,usenatbib]{mn2e}
\usepackage{graphicx}
\usepackage{ulem}
\usepackage{amsmath}
\title[Are there CBPs on the inclined orbits?]{Birth environment of circumbinary planets: are there CBPs on the inclined orbits?}
\author[Chuan-Tao Ma, Yan-Xiang Gong, et al.]
{Chuan-Tao Ma$^{1}$, Yan-Xiang Gong$^{1,2}$\thanks{yxgong@tsu.edu.cn}, Xiao-Mei Wu$^{1}$, Jianghui Ji$^{2}$\thanks{jijh@pmo.ac.cn}\\
$^{1}$College of Physics and Electronic Engineering, Taishan University, Taian 271000, China \\
$^{2}$CAS Key Laboratory of Planetary Sciences, Purple Mountain Observatory, Chinese Academy of Sciences, Nanjing 210008, China}
\begin{document}

\date{Received 2019 July 26; in original form 2019 September 1}

\pagerange{\pageref{firstpage}--\pageref{lastpage}} \pubyear{2019}

\maketitle

\label{firstpage}

\begin{abstract}
The inclination distribution of circumbinary planets (CBPs) is an important scientific issue.
It is of great significance in estimating the occurrence rate of CBPs and studying their formation and evolution.
Although the CBPs currently discovered by the transit method are nearly coplanar, the true inclination distribution of CBPs is still unknown. Previous researches on CBPs mostly regarded them as an isolated binary-planet system, without considering the birth environment of their host binaries. It is generally believed that almost all stars are born in clusters. Therefore, it is necessary to consider the impact of the close encounters of stars on the CBP systems. This article discusses how the close encounters of fly-by stars affect the inclination of CBP.
Based on extensive numerical simulations, we found that CBPs in close binary with a spacing of $\sim$ 0.2 au are almost unaffected by star fly-bys. Their orbits remain coplanar. However, when the spacing of the binary stars is greater than 1 au, the 2-3 fly-bys of the intruding star can excite a considerable inclination even for the CBP near the unstable boundary of the binary. For the planets in the outer region, a single star fly-by can excite an inclination to more than 5 degrees. Especially, CBPs in near polar or retrograde orbits can naturally form through bianry-star encounters.
If close binaries are born in open clusters, our simulations suggest that there may be high-inclination CBPs in binaries with spacing $>$ 1 au.
\end{abstract}


\section{Introduction}

Circumbinary planet (CBP) is one of the miraculous discoveries of exoplanet exploration. And up to now, more than 20 CBPs have been found (\citet{Schwarz2016}, \textit{http://www.univie.ac.at/adg/schwarz/multiple.html}). The most influential subset of them is the 11 planets detected by \textit{Kepler}, which constitute a small research sample. One remarkable feature of the orbits of \textit{Kepler}'s CBP is coplanarity, that is, the planes and their host binary are nearly in a same plane. The inclination angle $I$ between the two orbits is less than 2.5$^{\circ}$  \citep{Kostov2014}. This feature is partly due to the observational selection effect because that planets orbiting in the binary plane are more easily to be detected by the transit method. A very interesting question is: are there CBPs on the inclined orbits hidden in the universe? The inclination distribution of CBPs is an important scientific issue. For example, occurrence rates of CBP depend critically on the inclination distribution \citep{Armstrong2014}. If the underlying planetary inclination distribution is isotropic, the occurrence rate of CBP is significantly great than analogous rates for single stars. The prospects of finding planets transiting non-eclipsing binaries have been investigated by \citet{Martin2014}. They find there are potentially many of them lurking in the \textit{Kepler} photometric data. \citet{Zhang2019} provided a new tool for discovering potential polar CBPs, or misaligned CBPs of milder inclinations, from the existing ETV dataset of the \textit{Kepler}. Maybe CBPs on the inclined orbits will be detected in the near future.

Although no CBPs are found on the high-inclition orbits ($I>5^{\circ}$), some circumbinary disks (CBDs) on high inclination orbits have been discovered in recent years. For example, CBD in the IRS 43 system has an inclination $I>40^{\circ}$ \citep{Brinch2016, Czekala2019}. The CBD 99 Herculis and a planet-forming CBD in the young HD 98800 system are thought to have a polar configuration $I\sim90^{\circ}$ \citep{Kennedy2012, Kennedy2019}. Theoretical researches show that the evolution of a circumbinary discs tends to two extreme cases \citep{Brinch2016, Martin2017, Martin2018, Zanazzi2018, Lubow2018}. If the initial inclination angle of the disk is small, the result of the evolution is that the disk tends to be coplanar with the binary star. If the initial inclination angle of the disc is large ($>40^{\circ}$, in \citet{Martin2017}) and the eccentricity of the binary star is nonnegligible, the disc will evolve into a polar orbit. The presence of high-inclination discs indicates high-inclination CBPs may exist. However, above conclusions are based on the evolution of the isolated CBD systems themselves. Besides, only in the specific initial configurations can they evolve to the high-inclination orbits, such as a high initial inclination and a large eccentricity for the binary. Why the CBD initially has a large inclination still requires other mechanisms to explain. A noteworthy phenomenon is that the semi-major axis (SMA) of the binaries with high-inclination disks found now are relatively large (several tens of au), while the discs around close binaries are nearly coplanar \citep{Kennedy2012, Czekala2019}. For the \textit{Kepler} CBPs the SMA of their host binaries is small, about $0.1-0.2$ au. One possibility is that the discs around well-spaced binaries are susceptible to the surrounding environment, such as the close encounters with fly-by stars.

It is generally believed that most stars, and therefore most planetary systems, were born in clusters or associations \citep{Clarke2000, Lada2003, Pfalzner2013, Hao2013, Cai2017}. Some open clusters later dissolved, forming the current field stars. A notable example is that our own solar system may have formed in an open cluster. The chemical composition of objects in the Solar system, along with the orbital elements of Sedna, suggest that the Sun formed in a cluster with roughly $10^{3} \sim 10^{4}$ stars \citep{Adams2010}. Other work even suggests that the Sun was born in a massive cluster ($10^{4} \sim 10^{5}$ stars) because a several-thousand-star cluster is much too small to produce a nearby massive supernova \citep{Dukes2012}.
\citep{Malmberg2011} considered the effect of the close encounter between stars on the planetary systems in a single star system. They found that star fly-by is one of the possible mechanisms for explaining the generally high eccentricity of exoplanes (mainly discovered by RV methods). An interesting question is how does the cluster environment affect CBPs? In fact, the first CBP discovered, PSR B1620-26, is located in a globular cluster \citep{Sigurdsson2003}. According to the theory of star formation, it is impossible to form a close binary directly \citep{Moe2018}. One of the explanations is that they are formed by the interaction with other stars \citep{Martin2015, Hamers2016, Moe2018}, which means that they were in a star-dense environment before. Cluster provides such a possibility. If CBP was born in an open cluster, how do the star close encounters affect their orbital configuration? In particular, how do the star fly-bys affect the inclination distribution of CBP? It is what we focused in this work.

\section{Model and method}
We consider the effect of star fly-by on the inclination of CBP. The CBP family found by \textit{Kepler} is used as the reference to set the parameters of the binary star and the planet. The planet is initially on a coplanar and circular orbit. Its mass is 1 Saturn mass. The SMAs of the binaries in \textit{Kepler} CBP sample is about $a_{B}\sim$ 0.2 au. We take 0.2 au as the lower limit of $a_{B}$. According to \citet{Trilling2007}, the CBD around the close binaries with a SMA of 3 au is ubiquitous. 3 au is set as the upper limit of $a_{B}$. The mass ratio of the binary $m_{2}/m_{1}$ is 0.5. Here ${m_1}$ and ${m_2}$ are the mass of the primary and the secondary star, respectively. The eccentricity of the binary is $e_{B}=0.3$.

There is a stable boundary ($a_{c}$) around the binary beyond which the orbit of the planet is long-term stable (except for some unstable islands) \citep{Holman1999}.
\begin{equation}
\begin{array}{l}
{a_{c}} = \left[ {1.6 + 4.12\mu  + 5.1{e_B} - 4.27\mu {e_B} - } \right.\\
\left. {\quad \quad \,\;\;2.22e_B^2 - 5.09{\mu ^2} + 4.61{\mu ^2}e_B^2} \right]{a_B}
\end{array}
\end{equation}
where $\mu=m_{2}/(m_{1}+m_{2})$, $a_B$ and $e_B$ are the SMA and eccentricity of the binary, respectively. The SMA of CBP found by \textit{Kepler} is distributed between 1.1 $a_{c}$ and 7.4 $a_{c}$. Actually, most of them cluster at the vicinity of the unstable boundary of the binary ($\sim 1.1$ $a_{c}$) \citep{Gong2017}. It is generally believed that CBP formed in the outer region of the CBD and then migrated to their current location \citep{Thun2018}. \textit{Kepler}-1647b has the longest-period in the still-small family of CBPs ($\sim$ 1100 days, $a_{p}=$ 7.4 $a_{c}$) \citep{Kostov2016}, it may not have undergone significant migration. In this work we explore $a_{p}$ = 1.1, 4.1 and 7.1 $a_{c}$. Their values in au are shown in Table 1. The orbital phase angles of the planets and the intruding star are randomly and uniformly distributed between 0 and 360$^{\circ}$. Like most \textit{Kepler} CBP systems, we only consider single-planet system.

\begin{table}
% \centering
 \begin{minipage}{80mm}
  \caption{The semi-major axis (in au) of CBP that we considered. The results come from Equation (1). The eccentricity $e_{B}$ and mass ratio $\mu$ of binaries are 0.3 and 1/3, respectively.}
%  \begin{tabular}{@{}llrrrrlrlr@{}}
   \begin{tabular}{cccc}
  \hline
  $a_{B}$ (au)  & $a_{p}=1.1$ $a_{c}$ & $a_{p}=4.1$ $a_{c}$  &  $a_{p}=7.1$ $a_{c}$ \\
  \hline
   0.2  & 0.74  & 2.8 & 4.8  \\
   1.0  & 3.7  & 13.8 & 24  \\
   3.0  & 11  & 41 & 71  \\
\hline
\end{tabular}
\end{minipage}
\end{table}

We focus on stellar perturbers on parabolic orbits, which is common even for ONC-like clusters \citep{Olczak2010, Pfalzner2013, Xiang-Gruess2016}. It should be noted that in more dense clusters most stellar flybys are hyperbolic \citep{Spurzem2009, Cai2017}. We don't consider the hyperbolic orbit at present work for simplicity. The orbit of the stellar perturber is described by five parameters. They are the pericenter distance $q$, the inclination $i$, the longitude of ascending node $\Omega$, the argument of pericentre $\omega$, the true anomaly $f$, respectively \citep{Murray1999}. All the parameters are relative to the barycenter of the binary.

We use MERCURY\_RAS \citep{Smullen2016} for numerical integration. It is a modified version of MERCURY \citep{Chambers1999} that can be used to simulate a CBP system. The code has been well tested in our former work \citep{Gong2018}. We added a flyby star in the system. The parabolic orbits of the fly-by stars had been checked before the simulations. We found that although there is a negligible pulse in the eccentricity $e_{3}$ of the fly-by star near the pericenter, the intruding star is still in the near parabolic orbit $e_{3}\approx1$ after passing it. It means that the quadrupole moment of the binary has little effect on the orbit of the fly-by star in the cases we explored. In Figure 1, an example is shown. In our model, the fly-by star flies over the binary-planet system at an initial distance of 10,000 au from the barycenter of binary. When the star passes its pericenter and the distance between it and the binary exceeds 1000 au again, we record the orbital parameters of the planet.

\begin{figure*}
%  \vspace*{174pt}
\includegraphics[scale=0.6]{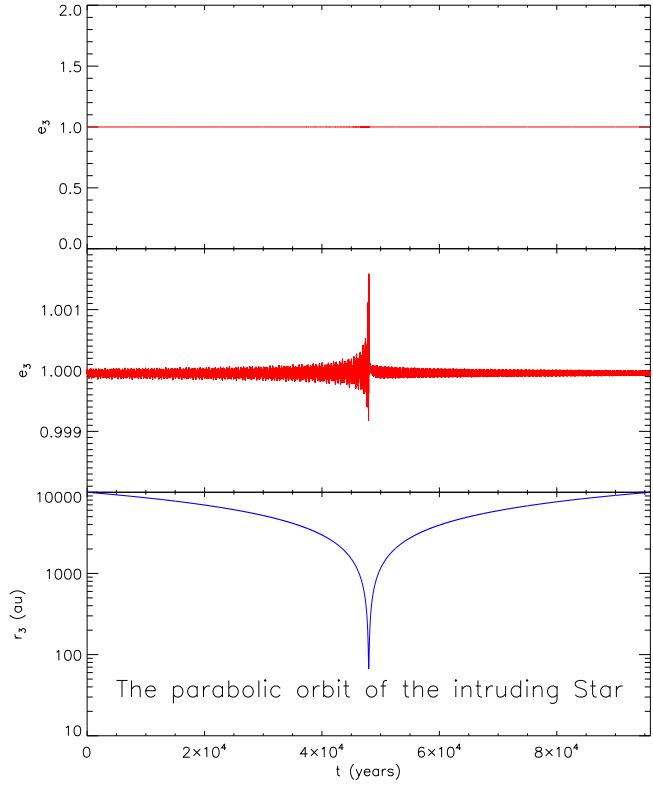}
  \centering
  \caption{The eccentricity evolution of the fly-by star. Top panel: the time evolution of the eccentricity of the fly-by star. Middle panel: zoom in on the top panel to show details. Bottom panel: the time evolution of the distance of the intruding star from the barycenter of the binary. The pericenter distance $q$ of the fly-by star is about 60 au. The semi-major axis of the binary is $a_{B}=$3 au.}\label{fig1}
 \end{figure*}

The long-term stability of the surviving planet is judged by the condition $q_{p}=a_{p}(1-e_{p})>a_{c}$ (the periastron of the planet is larger than $a_{c}$). Although the $a_{c}$ in \citet{Holman1999} was derived from the coplanar configuration, it still can applies to the non-coplanar configuration as shown in \citet{Pilat-Lohinger03}. They found that the inclination of the CBP has little effect on the stable boundary. Therefore, we still use the criterion in \citet{Holman1999} to check the orbital stability of the surviving CBPs.

As done in \citet{Malmberg2011}, we divided fly-bys into two different regimes, depending on $q$ : the strong regime ($q<$ 100 au) and the weak regime (100 au $< q <$ 1000 au). When the $q$ is very small, the binary may be disrupted. We discarded these cases when they occurred in the simulation. In other words, the criterion for us to judge whether a simulation is successful is that the close binary is still intact after star fly-bys.

The mass of the intruder star also has an effect on the simulation results. In the most simulations, we take $m_{3}=$1 $M_{\odot}$. In principle, a wide variety of stars with masses ranging from 0.08 $M_{\odot}$ up to 100 $M_{\odot}$ might act as a fly-by star in a cluster \citep{Pfalzner2018}. For example, it is generally believed that the solar system can closely encounter with a star with a mass of 25 $M_{\odot}$ \citep{Adams2010}. However, the more massive the stars, the fewer they are in a cluster. We just take $m_{3}=$ 20 $M_{\odot}$ as a representative of a high-mass perturber. The inclination of the fly-by star (relative to the binary plan) is randomly distributed between 0$^{\circ}$ and 180$^{\circ}$. All initial phase angles of the planets and the intruding star were assigned randomly and uniformly from 0 to 2$\pi$.

\section{Numerical results}

\subsection{A single fly-by}

\subsubsection{$q < 100$ au}

This close fly-by may occur in dense star clusters or in the inner part of an open cluster. \citet{Pfalzner2018} showed that a close fly-by with 50 au $< q <$ 100 au of a neighbouring star can reproduce the properties of the solar system. Numerical simulations showed in typical open clusters in the Solar neighbourhood containing hundreds or thousands of member stars, 10$\%$ to 20$\%$ of stars with mass $\geq1$ $M_{\odot}$ witness a fly-by $<$ 100 au \citep{Malmberg2007, Li2019}. In our simulation, the $q$ of the fly-by star is evenly and randomly distributed between 0 and 100 au, and the inclination angle is randomly between $0^{\circ}$ and $180^{\circ}$. The percentage of planets with an inclination greater than 5 degrees (PGT5) after a single star fly-by are shown in Table 2. Each statistical datum in Table 2 is based on 1000 realizations. The total number of the realizations is 72,000 in this work.

\begin{table}
% \centering
 \begin{minipage}{80mm}
  \caption{The percentage of planets with an inclination greater than 5 degrees (PGT5) after a single star fly-by. The SMA of the binary star is 0.2, 1 and 3 au, the eccentricity is 0.3. The masses are $m_{1}=$ 1 $M_{\odot}$ and $m_{2}=$ 0.5 $M_{\odot}$, respectively. The initial SMA of the planet is taken as 1.1, 4.1, 7. 1 $a_{c}$, here $a_{c}$ is the unstable boundary around the binary. The initial eccentricity of the planet is 0. We considered two types of fly-by stars, $m_{3}=1$ $M_{\odot}$ and $m_{3}=$ 20$M_{\odot}$. At the beginning of the simulation, the planet and the binary are coplanar. The pericenter distance of the fly-by star is uniformly and randomly distributed between 0 and 100 au. The lower limit of the $q$ satisfies that the intruding star doesn't destroy the binary system. The initial inclination of the fly-by star relative to the binary orbital plane is randomly distributed between 0$^{\circ}$ and 180$^{\circ}$. Each pair of data in Table 2 represents the PGT5 and the maximum inclination of the surviving planets. Each pair of data is a statistical result of 1000 realizations.}
%  \begin{tabular}{@{}llrrrrlrlr@{}}
   \begin{tabular}{cclll}
  \hline
  m$_3$ ($M_{\odot}$)  & $a_{B}$ (au)  & $a_{p}=1.1$ $a_{c}$ & $a_{p}=4.1$ $a_{c}$  &  $a_{p}=7.1$ $a_{c}$ \\
  \hline
  1           & 0.2  & 0.3$\%$, 18$^{\circ}$  & 2.0$\%$, 80$^{\circ}$ & 3.6$\%$, 75$^{\circ}$  \\
              & 1.0  & 1.1$\%$, 49$^{\circ}$  & 10.6$\%$, 119$^{\circ}$ & 21.5$\%$, 159$^{\circ}$  \\
              & 3.0  & 2.7$\%$, 35$^{\circ}$  & 39$\%$, 163$^{\circ}$ & 46$\%$, 173$^{\circ}$  \\
  \hline
  20          & 0.2  & 0.2$\%$, 13$^{\circ}$  &   2.1$\%$, 52$^{\circ}$ &   2.9$\%$, 125$^{\circ}$  \\
              & 1.0  & 0.7$\%$, 86$^{\circ}$  &   13$\%$, 164$^{\circ}$ &   26.8$\%$, 128$^{\circ}$  \\
              & 3.0  & 4.3$\%$, 84$^{\circ}$  &   32.7$\%$, 175$^{\circ}$  &   39.4$\%$, 177$^{\circ}$  \\
\hline
\end{tabular}
\end{minipage}
\end{table}

\begin{figure*}
%  \vspace*{174pt}
\includegraphics[scale=1.0]{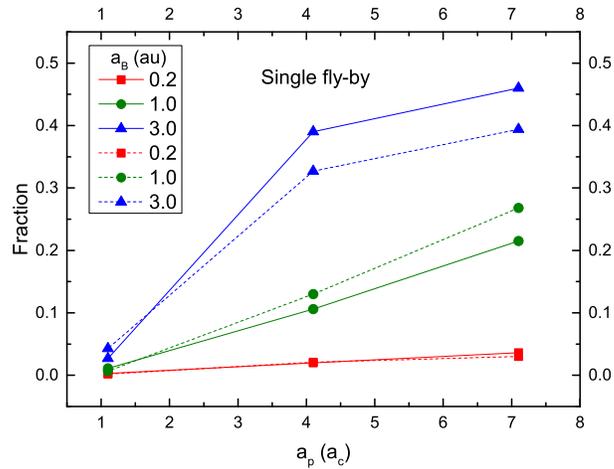}
  \centering
  \caption{The fraction of planets with an inclination greater than 5 degrees after a single star fly-by. Data are from Table 2. Red squares, green dots and blue triangles represent the data of $a_{B}=$ 0.2 au, 1 au and 3 au, respectively. Points connected by solid lines represent the case of $m_{3}=1$ $M_{\odot}$. Points connected by dashed lines represent the case of $m_{3}=20$ $M_{\odot}$.}\label{fig1}
 \end{figure*}

To clearly show the trend of PGT5 we plot them in Figure 2. It can be seen from Figure 2 that the percentage of planets with an inclination greater than 5$^{\circ}$ (PGT5) is gradually increasing as the $a_{B}$ increases (or see the same column in Table 2). The maximum inclination of CBPs is also gradually increasing, albeit with some statistical randomness. In the same binary system, the further the planet is, the easier the inclination is to be excited. That is, PGT5 gradually increases for $a_{p}$ from 1.1 to 7.1 $a_{c}$. A single fly-by can produce planets on retrograde orbits ($i_{p}>90^{\circ}$). For example, in the case of $a_{B}$ = 1 au, $a_{p}$ = 4.1 $a_{c} = 13.8$ au, the maximum inclination angle is 119$^{\circ}$. We also considered a massive intruder with $m_{3}=20$ $M_{\odot}$. On the whole, the PGT5 increases slightly. In Figure 3, we give an example of orbital evolution.

\begin{figure*}
%  \vspace*{174pt}
\includegraphics[scale=0.6]{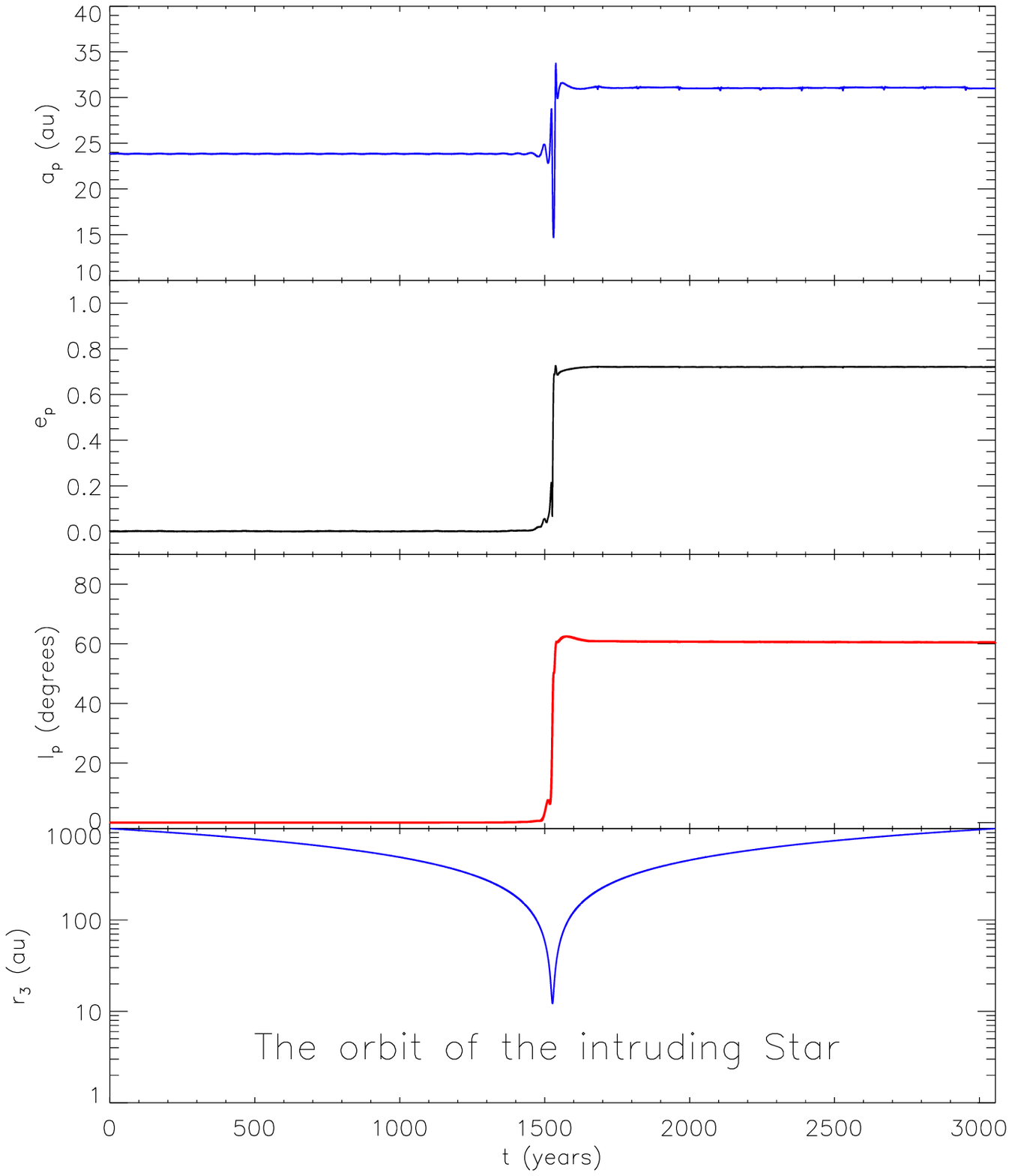}
  \centering
  \caption{Orbital evolution of a circumbinary planet. To show the details, we only simulate a piece of time before and after the fly-by. The time evolution of the SMA, eccentricity and inclination of the planet are shown in the first, second and third panel, respectively. The orbit of the intruding star is shown in the bottom panel. The parameters of the binary are: $m_{1}=M_{\odot}$, $m_{2}=0.5$ $M_{\odot}$, $a_{B}=1$ au, $e_{B}=0.3$. The planet is initially on a circular and coplanar orbit. Its SMA is $a_{p}=7.1$ $a_{c}\approx 24$ au. The intruding star is $m_{3}=1$ $M_{\odot}$ with the pericenter distance $q<100$ au. After star fly-by, the planet is excited to a high-inclination orbit ($i_{p} \simeq60^{\circ}$). Its orbit satisfies $q_{p}>a_{c}$ where $q_{p}=a_{p}(1-e_{p})=9.3$ au, $a_{c}=3.4$ au.} \label{fig1}
 \end{figure*}

Now we focus on the \textit{Kepler} CBPs. That is $a_{B}\sim$ 0.2 au, $a_{p}\sim1.1$ $a_{c}=0.74$ au. PGT5 is 0.3$\%$ for an intruder with $m_{3}=1$ $M_{\odot}$. Even at the location of 7.1 $a_{c}=4.8$ au from the binary, the percentage is still less than 5$\%$. For a massive flyby star $m_{3}=20$ $M_{\odot}$, the PGT5 does not change much. For \textit{Kepler}-like CBPs, the ratio is 0.2$\%$. The PGT5 of planets at 7.1 $a_{c}=4.8$ au is only 2.9$\%$. In fact, we also tested fly-by stars with a mass of 30 $M_{\odot}$, 50 $M_{\odot}$ and 100 $M_{\odot}$, the PGT5 is still $<1\%$. This is because the PGT5 mainly depends on the $q$ of the fly-by star. Only a small enough $q$ can cause significant orbit change to the inner planets. But for a fly-by star with a large mass, its $q$ cannot be too small because that the binary itself will disintegrate for this kind of close encounter. In summary, our calculations showed that for planets similar to those discovered by \textit{Kepler}, even if they were born in an open cluster, their inclination is almost unaffected by fly-by stars or, in other words, the surviving CBPs are nearly coplanar.

\subsubsection{100 au $< q <$ 1000 au}

When the $q$ of a fly-by star is greater than 100 au and less than 1000 au, a solar-mass intruder has little effect on the inclination of CBP (see Table 3 and Figure 4). Only planets with $a_{p}\geq4.1$ $a_{c}=41$ au in the binaries of $a_{B}=3$ au can be excited to an inclination of more than 5$^{\circ}$. In other cases, PGT5 is 0.

For fly-by stars with $m_{3}=20$ $M_{\odot}$, the PGT5 increases compared to $m_{3}=$ $M_{\odot}$ case. But for the binaries of $a_{B}\sim0.2$ au, PGT5 is still zero. The fly-by stars have little effect on the planetary systems. But for $a_{B}=3$ au, the orbit of the outer planet is easily excited to a large inclination. For example, for $a_{p}=4.1$ $a_{c}=41$ au, about one-fifth of the planets are excited to an inclination of more than 5$^{\circ}$. For $a_{p}=7.1$ $a_{c}=71$ au, PGT5 is $\sim$21$\%$ with a maximum inclination of 99$^{\circ}$.

\begin{table}
% \centering
 \begin{minipage}{80mm}
  \caption{PGT5. The pericenter distance of the fly-by star is uniformly and randomly distributed between 100 and 1000 au. Other parameters are the same as in Table 2.}
%  \begin{tabular}{@{}llrrrrlrlr@{}}
   \begin{tabular}{cclll}
  \hline
  m$_3$ ($M_{\odot}$)  & $a_{B}$ (au) & $a_{p}=1.1$ $a_{c}$ & $a_{p}=4.1$ $a_{c}$  &  $a_{p}=7.1$ $a_{c}$ \\
  \hline
  1     & 0.2  & 0$\%$, 0.005$^{\circ}$  & 0$\%$, 0.1$^{\circ}$ & 0$\%$, 0.25$^{\circ}$  \\
              & 1.0  & 0$\%$, 0.15$^{\circ}$  & 0$\%$, 1.3$^{\circ}$ & 0$\%$, 3$^{\circ}$  \\
              & 3.0  & 0$\%$, 0.68$^{\circ}$  & 0.5$\%$, 7.5$^{\circ}$ & 4.0$\%$, 35$^{\circ}$  \\
  \hline
  20 & 0.2   & 0$\%$, 0.01$^{\circ}$  &   0$\%$, 0.35$^{\circ}$ &   0$\%$, 0.8$^{\circ}$  \\
     & 1.0  & 0$\%$, 0.92$^{\circ}$  &   2.3$\%$, 7.1$^{\circ}$ &   9.5$\%$, 65$^{\circ}$  \\
     & 3.0  & 0.1$\%$, 5.6$^{\circ}$  &   18.1$\%$, 59$^{\circ}$ &   20.6$\%$, 99$^{\circ}$  \\
\hline
\end{tabular}
\end{minipage}
\end{table}

\begin{figure*}
%  \vspace*{174pt}
\includegraphics[scale=1.0]{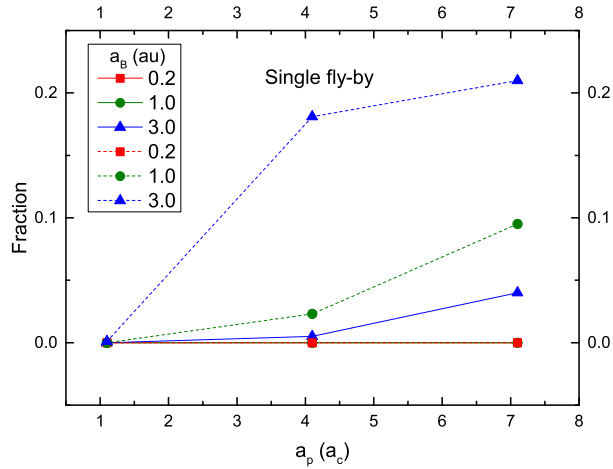}
  \centering
  \caption{Conventions are as in Figure 2. Data are from Table 3. The pericenter distance of the fly-by star is 100 au $< q <$ 1000 au.}\label{fig1}
 \end{figure*}

\subsection{Multiple fly-bys}

Above we only discuss a single fly-by. Multiple fly-bys also may occur in an open cluster. For example, \citet{Malmberg2011} showed the average number of fly-bys per sun-like star is 4 in the cluster with an initial number of stars of 700 and an initial half-mass radius of 0.38 pc. Besides, the more encounters a star undergoes, the smaller its fraction will be. For example, more than $60\%$ of the stars that have
undergone fly-bys in their reference cluster experience $\leq$ 5 fly-bys.  Close binaries ($a_{B} \leq 3$ au) that we consider in this work are known as `hard binaries' in cluster dynamics \citep{Malmberg2007}. They are more tightly bound and will not easily be broken up when they encounter another star. So the encounter rate of a close binary with a single star would be similar to the star-star encounter rate mentioned above.We explored the effect of 2-3 fly-bys on the orbits of CBP at present work. Certainly, during the dissolution of a long-lived cluster many more encounters exist. However, as mentioned above, their occurrence rate decreases significantly as the encounter number increases. On the other hand, with the dissolution of a cluster, the average encounter distance gets large. On average, despite the encounter number increases, the effects of fly-by stars on the planetary systems get weak. These issues complicate the problem and beyond the scope of our work.

We did the multiple fly-bys simulations as follows. After the first fly-by, we recorded the final position and velocity of the binary stars and planet. The data of the intruding star is discarded because it will has little effect on the orbit of the planet ($r_{3} \geq$ 1000 au). We started a new run. These data are used as the initial condition for binary and planet. In addition, we added a new random fly-by star in the system. As we done in the first fly-by, the new fly-by star flies over the binary-planet system at an initial distance of 10 000 au from the barycenter of binary. Similarly, the third fly-by was performed.

For the fly-by stars with $q<100$ au, the results of the three fly-bys are shown in Table 4. F1, F2 and F3 represent the results of the first, second, and third fly-by, respectively. For the case of $a_{B}=3.0$ au, after three fly-bys, most planets in the outer region are scattered out of the system. A few surviving planets can't give a meaningful statistical results. So we discarded these data. The fractions of planets with an inclination greater than 5 degrees after the first, second and third fly-by are given in Figure 5. In general, as the number of fly-bys increases, the PGT5 gets larger. The larger the $a_{B}$ and $a_{p}$, the larger the PGT5.

For the planetary system similar to \textit{Kepler} CBP, ie $a_{B}=0.2$ au, $a_{p}=1.1$ $a_{c}=0.74$ au in the Table 4, PGT5 is still less than $1\%$. After three fly-bys, PGT5 = $0.6\%$ and the maximum inclination is 19$^{\circ}$. Only for the planets in the outer region ($a_{p}=7.1$ $a_{c}=4.8$ au), more than $10\%$ of the planets are on the orbits greater than 5$^{\circ}$ after three fly-bys.

Table 5 and Figure 6 give the results of 100 au $< q <$ 1000 au. For the intruding stars of 1 $M_{\odot}$, the effect of multiple fly-bys on the planetary system is still limited. For the binary of $a_{B}=0.2$ au and 1 au, PGT5 is 0. Only for the systems with $a_{B}=3$ au and $a_{p}=7.1$ $a_{c}=71$ au, a considerable PGT5 is obtained after three fly-bys. For three successive fly-by, the results of PGT5 are 4$\%$, 10.4$\%$, and 15.9$\%$, respectively. The resulting maximum inclination is 39$^{\circ}$.

\begin{table}
% \centering
 \begin{minipage}{80mm}
  \caption{The result of multiple fly-bys. F1, F2, and F3 represent the results of the first, second, and third fly-by, respectively. For multiple fly-bys, we only consider the fly-by star of 1 $M_{\odot}$ mass. The pericenter distance of the fly-by star is $q < 100$ au as in Table 2.}
%  \begin{tabular}{@{}llrrrrlrlr@{}}
   \begin{tabular}{cclll}
  \hline
   $a_{B}$ (au) & Fly-by times  & $a_{p}=1.1$ $a_{c}$ & $a_{p}=4.1$ $a_{c}$  &  $a_{p}=7.1$ $a_{c}$ \\
  \hline
   0.2        &F1   & 0.3$\%$, 18$^{\circ}$  &   2$\%$, 80$^{\circ}$ &   3.6$\%$, 75$^{\circ}$  \\
              &F2    & 0.3$\%$, 29$^{\circ}$  &   6.1$\%$, 160$^{\circ}$ &   9.7$\%$, 169$^{\circ}$  \\
              &F3   & 0.6$\%$, 20.1$^{\circ}$  &   7.3$\%$, 150$^{\circ}$ &   16.4$\%$, 116$^{\circ}$  \\
  \hline
   1.0        &F1   & 1.1$\%$, 49$^{\circ}$  &   10.6$\%$, 119$^{\circ}$ &   21.5$\%$, 159$^{\circ}$  \\
              &F2   & 0.9$\%$, 26$^{\circ}$  &   23.6$\%$, 131$^{\circ}$ &   39.6$\%$, 171$^{\circ}$  \\
              &F3   & 1.4$\%$, 37$^{\circ}$  &   30.4$\%$, 144$^{\circ}$ &   57.2$\%$, 169$^{\circ}$  \\
  \hline
   3.0        &F1   & 2.7$\%$, 35$^{\circ}$  &   39$\%$, 163$^{\circ}$ &   46$\%$, 173$^{\circ}$  \\
              &F2   & 4.5$\%$, 18.2$^{\circ}$  &   36$\%$, 165$^{\circ}$  &  ---   \\
              &F3   & 9.7$\%$, 56$^{\circ}$  &   --- &   ---  \\
\hline
\end{tabular}
\end{minipage}
\end{table}

\begin{figure*}
%  \vspace*{174pt}
\includegraphics[scale=1.0]{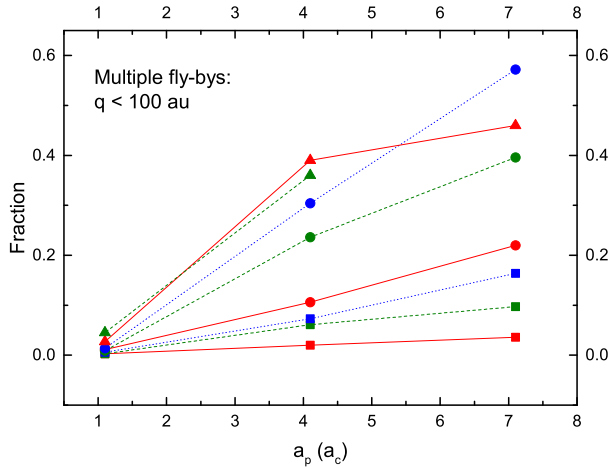}
  \centering
  \caption{The fraction of planets with an inclination greater than 5 degrees after the first, second and third fly-by. Data are from Table 4. The squares, circles and triangles represent the results of $a_{B}=$ 0.2 au, 1 au and 3 au, respectively. The results of the first, second and third fly-by are shown in red, green and blue, respectively. The pericenter distance of the fly-by star is $q < 100$ au.}\label{fig1}
 \end{figure*}

\begin{table}
% \centering
 \begin{minipage}{80mm}
  \caption{The result of multiple fly-bys. The pericenter distance of the fly-by star is uniformly and randomly distributed between 100 au and 1000 au. Other parameters are the same as in Table 4.}
%  \begin{tabular}{@{}llrrrrlrlr@{}}
   \begin{tabular}{cclll}
  \hline
   $a_{B}$ (au) & Fly-by times  & $a_{p}=1.1$ $a_{c}$ & $a_{p}=4.1$ $a_{c}$  &  $a_{p}=7.1$ $a_{c}$ \\
  \hline
   0.2        &F1   & 0$\%$, 0.005$^{\circ}$  &   0$\%$, 0.105$^{\circ}$ &   0$\%$, 0.25$^{\circ}$  \\
              &F2   & 0$\%$, 0.004$^{\circ}$  &   0$\%$, 0.14$^{\circ}$ &    0$\%$, 0.31$^{\circ}$  \\
              &F3   & 0$\%$, 0.005$^{\circ}$  &   0$\%$, 0.15$^{\circ}$ &    0$\%$, 0.32$^{\circ}$  \\
  \hline
   1.0        &F1   & 0$\%$, 0.15$^{\circ}$  &   0$\%$, 1.32$^{\circ}$ &   0$\%$, 3$^{\circ}$  \\
              &F2   & 0$\%$, 0.18$^{\circ}$  &   0$\%$, 1.35$^{\circ}$ &    0$\%$, 3.3$^{\circ}$  \\
              &F3   & 0$\%$, 0.19$^{\circ}$  &    0$\%$, 1.31$^{\circ}$ &     0$\%$, 2.9$^{\circ}$  \\
  \hline
   3.0        &F1   & 0$\%$, 0.68$^{\circ}$  &   0.5$\%$, 7.5$^{\circ}$ &   4$\%$, 35$^{\circ}$  \\
              &F2   & 0$\%$, 0.74$^{\circ}$  &   1$\%$, 8.5$^{\circ}$ &   10$\%$, 36$^{\circ}$  \\
              &F3   & 0$\%$, 0.93$^{\circ}$  &   2.2$\%$, 7.7$^{\circ}$ &   16$\%$, 39$^{\circ}$  \\
\hline
\end{tabular}
\end{minipage}
\end{table}

\begin{figure*}
%  \vspace*{174pt}
\includegraphics[scale=1.0]{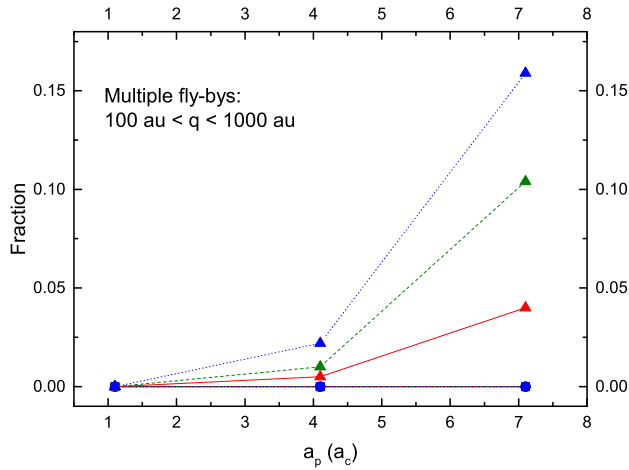}
  \centering
  \caption{Conventions are as in Figure 5. Data are from Table 5. The pericenter distance of the fly-by star is 100 au $< q <$ 1000 au.}\label{fig1}
 \end{figure*}

\section{Summary}

It is generally believed that most stars are born in a clusters. The nascent planetary system of these stars is vulnerable to the perturbations of other stars. This article considers the effects of star fly-by(s) on the inclination of CBP. Our main conclusions are as follows.

The CBP systems with $a_{B}\sim$ 0.2 au similar to those discovered by \textit{Kepler} are almost unaffected by a single star fly-by. Their orbits remain coplanar. Even a relatively close flyby, or a massive intruder, has little effect on the inclination of the planet. Our simulations also showed that several successive fly-bys do not cause significant inclination change for CBP in tight binaries. These results imply that for planets similar to those discovered by \textit{Kepler}, even if they were born in an open cluster, their orbits keep nearly coplanar.

However, for binaries with $a_{B}>$ 1 au, the planets in the outer region ($a_{p}>4.1a_{c}$) will be affected by the stellar fly-by(s). A single fly-by can excite the inclination of CBP to more than 5$^{\circ}$ (PGT5 $>$ 10$\%$). For planets close to the unstable boundary of the binary, the 2-3 close fly-bys cause significant inclination excitation. If, like most stars, close binaries are born in clusters, our simulations suggested that there may be high-inclination planets around binaries with $a_{B}>$ 1 au. Besides, it is worth mentioning that CBPs in near polar or retrograde orbits can naturally form in clusters through this mechanism.

At present, CBDs with high inclination angles are found in binaries with a large $a_{B}$, while CBDs (and CBPs) found in close binaries are almost coplanar. Our research implies that the star fly-by may be one of the possible mechanisms accounting for this difference. If the nascent CBD and their host binaries are coplanar, this difference can be explained by the fact that the CBD in the binaries with a larger $a_{B}$ is more susceptible to stellar fly-bys. Generation of highly inclined protoplanetary discs in single star systems through a single stellar fly-by has been explored in \citet{Xiang-Gruess2016}. We will discuss how it works for CBD in the near future.

An in-depth exploration of \textit{Kepler}'s existing data, as well as the ongoing PLATO and TESS mission, is expected to find more CBPs. The orbital inclination distribution of CBP will be the key information to understand their formation and evolution.

\section*{Acknowledgments}

We thank the reviewer for his/her valuable and insightful comments, which have improved our manuscript substantially. This work is financially supported by National Natural Science Foundation of China (Grants No. 11573018, 11773081, 11573018), CAS Interdisciplinary Innovation Team, and the Foundation of Minor Planets of Purple Mountain Observatory. Ma Ch.-T. also acknowledges the support from Scientific Research Projects for Universities in Shandong Province, China (J18KB101). Thanks to Jun-Yi Che, Shuang Liu, Yuan Yang, Zheng-Yi Xu and Bing-Xin Yan for their help in numerical calculations.

%%%%%%%%%%%%%%%%%%%%%%%%%%%%%%%%%%%%%%%%%%%%%%%%%%

%%%%%%%%%%%%%%%%% APPENDICES %%%%%%%%%%%%%%%%%%%%%

%%%%%%%%%%%%%%%%%%%%%%%%%%%%%%%%%%%%%%%%%%%%%%%%%%

\label{lastpage}

\end{document}